%% file: main.tex
\documentclass[letterpaper,twocolumn,10pt]{article}
\usepackage{usenix,epsfig,endnotes}
\usepackage{tugraz_defaults}

\input{numbers.tex}
\begin{document}


\date{}

\title{Meltdown}

\author{
 {\rm Moritz Lipp}$^1$, {\rm Michael Schwarz}$^1$, {\rm Daniel Gruss}$^1$, {\rm Thomas Prescher}$^2$,  {\rm Werner Haas}$^2$, \\ {\rm Stefan Mangard}$^1$, {\rm Paul Kocher}$^3$, {\rm Daniel Genkin}$^4$, {\rm Yuval Yarom}$^5$, {\rm Mike Hamburg}$^6$ \\
$^1$ Graz University of Technology \\
$^2$ Cyberus Technology GmbH \\
$^3$ Independent \\
$^4$ University of Pennsylvania and University of Maryland \\
$^5$ University of Adelaide and Data61 \\
$^6$ Rambus, Cryptography Research Division
}

\maketitle



\newcommand{\AttackName}{Meltdown\xspace} 
\newcommand{\DefenseName}{KAISER\xspace}
\newcommand{\Instructionname}{Transient\xspace}
\newcommand{\instructionname}{transient\xspace}



\subsection*{Abstract}
The security of computer systems fundamentally relies on memory isolation, \eg kernel address ranges are marked as non-accessible and are protected from user access.
In this paper, we present \AttackName.
\AttackName exploits side effects of out-of-order execution on modern processors to read arbitrary kernel-memory locations including personal data and passwords.
Out-of-order execution is an indispensable performance feature and present in a wide range of modern processors.
The attack works on different Intel microarchitectures since at least 2010 and potentially other processors are affected.
The root cause of \AttackName is the hardware.
The attack is independent of the operating system, and it does not rely on any software vulnerabilities.
\AttackName breaks all security assumptions given by address space isolation as well as paravirtualized environments and, thus, every security mechanism building upon this foundation.
On affected systems, \AttackName enables an adversary to read memory of other processes or virtual machines in the cloud without any permissions or privileges, affecting millions of customers and virtually every user of a personal computer.
We show that the \DefenseName defense mechanism for KASLR~\cite{Gruss2017Kaslr} has the important (but inadvertent) side effect of impeding \AttackName.
We stress that \DefenseName must be deployed immediately to prevent large-scale exploitation of this severe information leakage.



\section{Introduction} 
One of the central security features of today's operating systems is memory isolation.
Operating systems ensure that user applications cannot access each other's memories and prevent user applications from reading or writing kernel memory.
This isolation is a cornerstone of our computing environments and allows running multiple applications on personal devices or executing processes of multiple users on a single machine in the cloud.

On modern processors, the isolation between the kernel and user processes is typically realized by a supervisor bit of the processor that defines whether a memory page of the kernel can be accessed or not.
The basic idea is that this bit can only be set when entering kernel code and it is cleared when switching to user processes.
This hardware feature allows operating systems to map the kernel into the address space of every process and to have very efficient transitions from the user process to the kernel, \eg for interrupt handling.
Consequently, in practice, there is no change of the memory mapping when switching from a user process to the kernel.


In this work, we present \AttackName\footnote{This attack was independently found by the authors of this paper and Jann Horn from Google Project Zero.}.
\AttackName is a novel attack that allows overcoming memory isolation completely by providing a simple way for any user process to read the entire kernel memory of the machine it executes on, including all physical memory mapped in the kernel region.
\AttackName does not exploit any software vulnerability, \ie it works on all major operating systems.
Instead, \AttackName exploits side-channel information available on most modern processors, \eg modern Intel microarchitectures since 2010 and possibly on other CPUs of other vendors.

While side-channel attacks typically require very specific knowledge about the target application and only leak information about secrets of the target application, \AttackName allows an adversary who can run code on the vulnerable processor to easily dump the entire kernel address space, including any mapped physical memory.
The root cause of the simplicity and strength of \AttackName are side effects caused by \emph{out-of-order execution}.

Out-of-order execution is an important performance feature of today's processors in order to overcome latencies of busy execution units, \eg a memory fetch unit needs to wait for data arrival from memory.
Instead of stalling the execution, modern processors run operations \emph{out-of-order} \ie they look ahead and schedule subsequent operations to idle execution units of the processor.
However, such operations often have unwanted side-effects, \eg timing differences~\cite{Osvik2006,Yarom2014,Gruss2015Template} can leak information from both sequential and out-of-order execution.

From a security perspective, one observation is particularly significant:
Out-of-order; vulnerable CPUs allow an unprivileged process to load data from a privileged (kernel or physical) address into a temporary CPU register.
Moreover, the CPU even performs further computations based on this register value, \eg access to an array based on the register value.
The processor ensures correct program execution, by simply discarding the results of the memory lookups (\eg the modified register states), if it turns out that an instruction should not have been executed.
Hence, on the architectural level (\eg the abstract definition of how the processor should perform computations), no security problem arises.

However, we observed that out-of-order memory lookups influence the cache, which in turn can be detected through the cache side channel.
As a result, an attacker can dump the entire kernel memory by reading privileged memory in an out-of-order execution stream, and transmit the data from this elusive state via a microarchitectural covert channel (\eg \FlushReload) to the outside world.
On the receiving end of the covert channel, the register value is reconstructed.
Hence, on the microarchitectural level (\eg the actual hardware implementation), there is an exploitable security problem.

\AttackName breaks all security assumptions given by the CPU's memory isolation capabilities.
We evaluated the attack on modern desktop machines and laptops, as well as servers in the cloud.
\AttackName allows an unprivileged process to read data mapped in the kernel address space, including the entire physical memory on Linux and OS X, and a large fraction of the physical memory on Windows.
This may include physical memory of other processes, the kernel, and in case of kernel-sharing sandbox solutions (\eg Docker, LXC) or Xen in paravirtualization mode, memory of the kernel (or hypervisor), and other co-located instances.
While the performance heavily depends on the specific machine, \eg processor speed, TLB and cache sizes, and DRAM speed, we can dump kernel and physical memory with up to \SI{503}{KB/\second}.
Hence, an enormous number of systems are affected.

The countermeasure \DefenseName~\cite{Gruss2017Kaslr}, originally developed to prevent side-channel attacks targeting KASLR, inadvertently protects against \AttackName as well.
Our evaluation shows that \DefenseName prevents \AttackName to a large extent.
Consequently, we stress that it is of utmost importance to deploy \DefenseName on all operating systems immediately.
Fortunately, during a responsible disclosure window, the three major operating systems (Windows, Linux, and OS X) implemented variants of \DefenseName and will roll out these patches in the near future.

\AttackName is distinct from the Spectre Attacks~\cite{Kocher2017} in several ways, notably that Spectre requires tailoring to the victim process's software environment, but applies more broadly to CPUs and is not mitigated by \DefenseName.

\paragraph{Contributions.}
The contributions of this work are:
\begin{compactenum}
\item We describe out-of-order execution as a new, extremely powerful, software-based side channel.
\item We show how out-of-order execution can be combined with a microarchitectural covert channel to transfer the data from an elusive state to a receiver on the outside.
\item We present an end-to-end attack combining out-of-order execution with exception handlers or TSX, to read arbitrary physical memory without any permissions or privileges, on laptops, desktop machines, and on public cloud machines.
\item We evaluate the performance of \AttackName and the effects of \DefenseName on it.
\end{compactenum}

\paragraph{Outline.} The remainder of this paper is structured as follows:
In \cref{sec:background}, we describe the fundamental problem which is introduced with out-of-order execution.
In \cref{sec:toy-example}, we provide a toy example illustrating the side channel \AttackName exploits.
In \cref{sec:building-blocks}, we describe the building blocks of the full \AttackName attack.
In \cref{sec:meltdown}, we present the \AttackName attack.
In \cref{sec:evaluation}, we evaluate the performance of the \AttackName attack on several different systems.
In \cref{sec:countermeasures}, we discuss the effects of the software-based \DefenseName countermeasure and propose solutions in hardware.
In \cref{sec:discussion}, we discuss related work and conclude our work in \cref{sec:conclusion}.


\section{Background}
\label{sec:background}

In this section, we provide background on out-of-order execution, address translation, and cache attacks.

\subsection{Out-of-order execution}\label{sec:background:spex}

Out-of-order execution is an optimization technique that allows to maximize the utilization of all execution units of a CPU core as exhaustive as possible.
Instead of processing instructions strictly in the sequential program order, the CPU executes them as soon as all required resources are available.
While the execution unit of the current operation is occupied, other execution units can run ahead.
Hence, instructions can be run in parallel as long as their results follow the architectural definition.

In practice, CPUs supporting out-of-order execution support running operations \emph{speculatively} to the extent that the processor's out-of-order logic processes instructions before the CPU is certain whether the instruction will be needed and committed.
In this paper, we refer to speculative execution in a more restricted meaning, where it refers to an instruction sequence following a branch, and use the term out-of-order execution to refer to any way of getting an operation executed before the processor has committed the results of all prior instructions.

In 1967, Tomasulo~\cite{Tomasulo1967} developed an algorithm~\cite{Tomasulo1967} that enabled dynamic scheduling of instructions to allow out-of-order execution.
Tomasulo~\cite{Tomasulo1967} introduced a unified reservation station that allows a CPU to use a data value as it has been computed instead of storing it to a register and re-reading it.
The reservation station renames registers to allow instructions that operate on the same physical registers to  use the last logical one to solve read-after-write (RAW), write-after-read (WAR) and write-after-write (WAW) hazards.
Furthermore, the reservation unit connects all execution units via a common data bus (CDB).
If an operand is not available, the reservation unit can listen on the CDB until it is available and then directly begin the execution of the instruction.

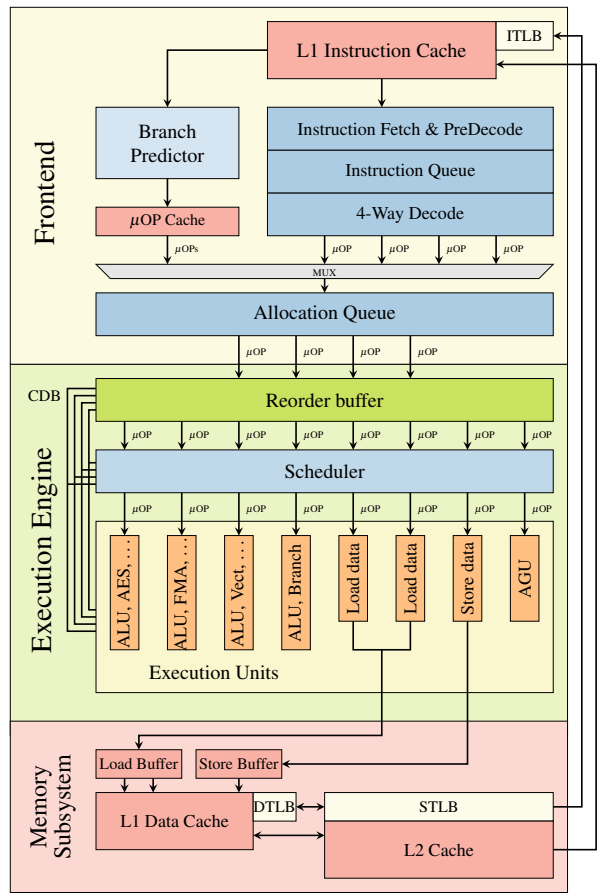
\begin{figure}[t]
 \centering
 \tikzsetnextfilename{core-skylake}
 \resizebox{\hsize}{!}{%
 \input{images/core-skylake.tikz}
 }%
 \caption{Simplified illustration of a single core of the Intel's Skylake microarchitecture. Instructions are decoded into \muops and executed out-of-order in the execution engine by individual execution units.}
 \label{fig:core-skylake}
\end{figure}

On the Intel architecture, the pipeline consists of the front-end, the execution engine (back-end) and the memory subsystem~\cite{Intel_opt}.
x86 instructions are fetched by the front-end from the memory and decoded to micro-operations (\muops) which are continuously sent to the execution engine.
Out-of-order execution is implemented within the execution engine as illustrated in~\cref{fig:core-skylake}.
The \textit{Reorder Buffer} is responsible for register allocation, register renaming and retiring.
Additionally, other optimizations like move elimination or the recognition of zeroing idioms are directly handled by the reorder buffer.
The \muops are forwarded to the \textit{Unified Reservation Station} that queues the operations on exit ports that are connected to \textit{Execution Units}.
Each execution unit can perform different tasks like ALU operations, AES operations, address generation units (AGU) or memory loads and stores.
AGUs as well as load and store execution units are directly connected to the memory subsystem to process its requests.

Since CPUs usually do not run linear instruction streams, they have branch prediction units that are used to obtain an educated guess of which instruction will be executed next.
Branch predictors try to determine which direction of a branch will be taken before its condition is actually evaluated.
Instructions that lie on that path and do not have any dependencies can be executed in advance and their results immediately used if the prediction was correct.
If the prediction was incorrect, the reorder buffer allows to rollback by clearing the reorder buffer and re-initializing the unified reservation station.

Various approaches to predict the branch exist:
With static branch prediction~\cite{Hennessy2011}, the outcome of the branch is solely based on the instruction itself.
Dynamic branch prediction~\cite{Cheng2000} gathers statistics at run-time to predict the outcome.
One-level branch prediction uses a 1-bit or 2-bit counter to record the last outcome of the branch~\cite{LeeDynamicBranchPrediction}.
Modern processors often use two-level adaptive predictors~\cite{Yeh1991} that remember the history of the last $n$ outcomes allow to predict regularly recurring patterns.
More recently, ideas to use neural branch prediction~\cite{Vintan1999,Jimenez2001,Teran2016} have been picked up and integrated into CPU architectures~\cite{AMDRyzen}.

\subsection{Address Spaces}\label{sec:background:vaddr}
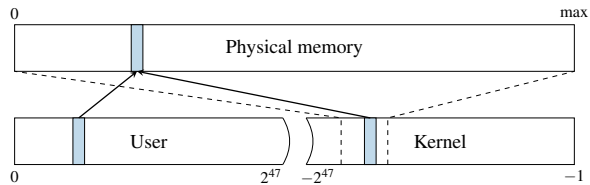
\begin{figure}[t]
 \centering
 \tikzsetnextfilename{identity-mapping}
 \resizebox{\hsize}{!}{%
 \input{images/identity-mapping.tikz}
 }%
 \caption{The physical memory is directly mapped in the kernel at a certain offset. A physical address (blue) which is mapped accessible for the user space is also mapped in the kernel space through the direct mapping.}
 \label{fig:identity-mapping}
\end{figure}

To isolate processes from each other, CPUs support virtual address spaces where virtual addresses are translated to physical addresses.
A virtual address space is divided into a set of pages that can be individually mapped to physical memory through a multi-level page translation table.
The translation tables define the actual virtual to physical mapping and also protection properties that are used to enforce privilege checks, such as readable, writable, executable and user-accessible.
The currently used translation table that is held in a special CPU register.
On each context switch, the operating system updates this register with the next process' translation table address in order to implement per process virtual address spaces.
Because of that, each process can only reference data that belongs to its own virtual address space.
Each virtual address space itself is split into a user and a kernel part.
While the user address space can be accessed by the running application, the kernel address space can only be accessed if the CPU is running in privileged mode.
This is enforced by the operating system disabling the user-accessible property of the corresponding translation tables.
The kernel address space does not only have memory mapped for the kernel's own usage, but it also needs to perform operations on user pages, \eg filling them with data.
Consequently, the entire physical memory is typically mapped in the kernel.
On Linux and OS X, this is done via a direct-physical map, \ie the entire physical memory is directly mapped to a pre-defined virtual address (\cf \cref{fig:identity-mapping}).

Instead of a direct-physical map, Windows maintains a multiple so-called \emph{paged pools}, \emph{non-paged pools}, and the \emph{system cache}.
These pools are virtual memory regions in the kernel address space mapping physical pages to virtual addresses which are either required to remain in the memory (non-paged pool) or can be removed from the memory because a copy is already stored on the disk (paged pool).
The \emph{system cache} further contains mappings of all file-backed pages.
Combined, these memory pools will typically map a large fraction of the physical memory into the kernel address space of every process.

The exploitation of memory corruption bugs often requires the knowledge of addresses of specific data.
In order to impede such attacks, address space layout randomization (ASLR) has been introduced as well as non-executable stacks and stack canaries.
In order to protect the kernel, KASLR randomizes the offsets where drivers are located on every boot, making attacks harder as they now require to guess the location of kernel data structures.
However, side-channel attacks allow to detect the exact location of kernel data structures~\cite{Gruss2016Prefetch,Hund2013,Jang2016} or derandomize ASLR in JavaScript~\cite{Gras2017AnC}.
A combination of a software bug and the knowledge of these addresses can lead to privileged code execution.

\subsection{Cache Attacks}\label{sec:background:caches}
In order to speed-up memory accesses and address translation, the CPU contains small memory buffers, called caches, that store frequently used data.
CPU caches hide slow memory access latencies by buffering frequently used data in smaller and faster internal memory.
Modern CPUs have multiple levels of caches that are either private to its cores or shared among them.
Address space translation tables are also stored in memory and are also cached in the regular caches.

Cache side-channel attacks exploit timing differences that are introduced by the caches.
Different cache attack techniques have been proposed and demonstrated in the past, including \EvictTime~\cite{Osvik2006}, \PrimeProbe~\cite{Osvik2006,Percival2005}, and \FlushReload~\cite{Yarom2014}.
\FlushReload attacks work on a single cache line granularity.
These attacks exploit the shared, inclusive last-level cache.
An attacker frequently flushes a targeted memory location using the \clflush instruction.
By measuring the time it takes to reload the data, the attacker determines whether data was loaded into the cache by another process in the meantime.
The \FlushReload attack has been used for attacks on various computations, \eg cryptographic algorithms~\cite{Yarom2014,Irazoqui2014,Benger2014}, web server function calls~\cite{Zhang2014}, user input~\cite{Gruss2015Template,Lipp2016,Schwarz2018KeyDrown}, and kernel addressing information~\cite{Gruss2016Prefetch}.

A special use case are covert channels.
Here the attacker controls both, the part that induces the side effect, and the part that measures the side effect.
This can be used to leak information from one security domain to another, while bypassing any boundaries existing on the architectural level or above.
Both \PrimeProbe and \FlushReload have been used in high-performance covert channels~\cite{Liu2015,Maurice2017Hello,Gruss2016Flush}.


\section{A Toy Example}
\label{sec:toy-example}

In this section, we start with a toy example, a simple code snippet, to illustrate that out-of-order execution can change the microarchitectural state in a way that leaks information.
However, despite its simplicity, it is used as a basis for \Cref{sec:building-blocks} and \Cref{sec:meltdown}, where we show how this change in state can be exploited for an attack.

\begin{lstlisting}[language=C,style=customc,float,caption={A toy example to illustrate side-effects of out-of-order execution.},label={lst:toy-example}]
raise_exception();
// the line below is never reached
access(probe_array[data * 4096]);
\end{lstlisting}

\Cref{lst:toy-example} shows a simple code snippet first raising an (unhandled) exception and then accessing an array.
The property of an exception is that the control flow does not continue with the code after the exception, but jumps to an exception handler in the operating system.
Regardless of whether this exception is raised due to a memory access, \eg by accessing an invalid address, or due to any other CPU exception, \eg a division by zero, the control flow continues in the kernel and not with the next user space instruction.

\begin{figure}[t]
 \centering
\tikzsetnextfilename{toy-illustration}
\resizebox{!}{3.5cm}{
\input{images/toy-illustration.tikz}
}
 \caption{If an executed instruction causes an exception, diverting the control flow to an exception handler, the subsequent instruction must not be executed anymore. Due to out-of-order execution, the subsequent instructions may already have been partially executed, but not retired. However, the architectural effects of the execution will be discarded.}
 \label{fig:toy-illustration}
\end{figure}
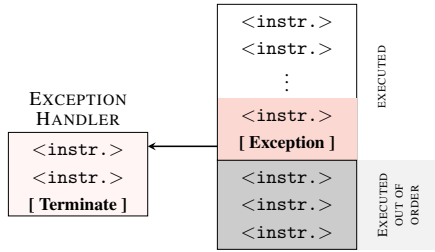

Thus, our toy example cannot access the array in theory, as the exception immediately traps to the kernel and terminates the application.
However, due to the out-of-order execution, the CPU might have already executed the following instructions as there is no dependency on the exception.
This is illustrated in \Cref{fig:toy-illustration}.
Due to the exception, the instructions executed out of order are not retired and, thus, never have architectural effects.

Although the instructions executed out of order do not have any visible architectural effect on registers or memory, they have microarchitectural side effects.
During the out-of-order execution, the referenced memory is fetched into a register and is also stored in the cache.
If the out-of-order execution has to be discarded, the register and memory contents are never committed.
Nevertheless, the cached memory contents are kept in the cache.
We can leverage a microarchitectural side-channel attack such as \FlushReload~\cite{Yarom2014}, which detects whether a specific memory location is cached, to make this microarchitectural state visible.
There are other side channels as well which also detect whether a specific memory location is cached, including \PrimeProbe~\cite{Osvik2006,Liu2015,Maurice2017Hello}, \EvictReload~\cite{Lipp2016}, or \FlushFlush~\cite{Gruss2016Flush}.
However, as \FlushReload is the most accurate known cache side channel and is simple to implement, we do not consider any other side channel for this example.

Based on the value of \texttt{data} in this toy example, a different part of the cache is accessed when executing the memory access out of order.
As \texttt{data} is multiplied by \SIx{4096}, data accesses to \texttt{probe\_array} are scattered over the array with a distance of \SI{4}{\kilo B} (assuming an \SI{1}{B} data type for \texttt{probe\_array}).
Thus, there is an injective mapping from the value of \texttt{data} to a memory page, \ie there are no two different values of data which result in an access to the same page.
Consequently, if a cache line of a page is cached, we know the value of \texttt{data}.
The spreading over different pages eliminates false positives due to the prefetcher, as the prefetcher cannot access data across page boundaries~\cite{Intel_opt}.

\begin{figure}[t]
 \centering
\tikzsetnextfilename{fr-toy}
\input{images/fr-toy.tikz}
 \caption{Even if a memory location is only accessed during out-of-order execution, it remains cached. Iterating over the \SIx{256} pages of \texttt{probe\_array} shows one cache hit, exactly on the page that was accessed during the out-of-order execution.}
 \label{fig:fr-toy-example}
\end{figure}
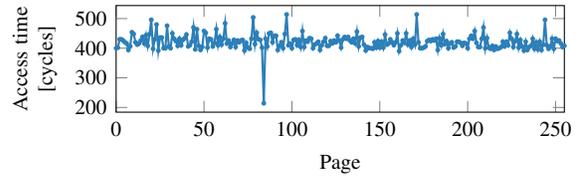

\Cref{fig:fr-toy-example} shows the result of a \FlushReload measurement iterating over all pages, after executing the out-of-order snippet with \texttt{data = 84}.
Although the array access should not have happened due to the exception, we can clearly see that the index which would have been accessed is cached.
Iterating over all pages (\eg in the exception handler) shows only a cache hit for page \SI{84}.
This shows that even instructions which are never actually executed, change the microarchitectural state of the CPU.
\Cref{sec:building-blocks} modifies this toy example to not read a value, but to leak an inaccessible secret.


\section{Building Blocks of the Attack}
\label{sec:building-blocks}

The toy example in~\cref{sec:toy-example} illustrated that side-effects of out-of-order execution can modify the microarchitectural state to leak information.
While the code snippet reveals the data value passed to a cache-side channel, we want to show how this technique can be leveraged to leak otherwise inaccessible secrets.
In this section, we want to generalize and discuss the necessary building blocks to exploit out-of-order execution for an attack.

The adversary targets a secret value that is kept somewhere in physical memory.
Note that register contents are also stored in memory upon context switches, \ie they are also stored in physical memory.
As described in \cref{sec:background:vaddr}, the address space of every process typically includes the entire user space, as well as the entire kernel space, which typically also has all physical memory (in-use) mapped.
However, these memory regions are only accessible in privileged mode (\cf \cref{sec:background:vaddr}).

In this work, we demonstrate leaking secrets by bypassing the privileged-mode isolation, giving an attacker full read access to the entire kernel space including any physical memory mapped, including the physical memory of any other process and the kernel.
Note that Kocher~\etal\cite{Kocher2017} pursue an orthogonal approach, called Spectre Attacks, which trick speculative executed instructions into leaking information that the victim process is authorized to access.
As a result, Spectre Attacks lack the privilege escalation aspect of \AttackName and require tailoring to the victim process's software environment, but apply more broadly to CPUs that support speculative execution and are not stopped by \DefenseName.

\begin{figure}[t]
 \centering
\tikzsetnextfilename{building-blocks}
\input{images/building-blocks.tikz}
 \caption{The \AttackName attack uses exception handling or suppression, \eg TSX, to run a series of transient instructions. These transient instructions obtain a (persistent) secret value and change the microarchitectural state of the processor based on this secret value. This forms the sending part of a microarchitectural covert channel. The receiving side reads the microarchitectural state, making it architectural and recovering the secret value.}
 \label{fig:building-blocks}
\end{figure}
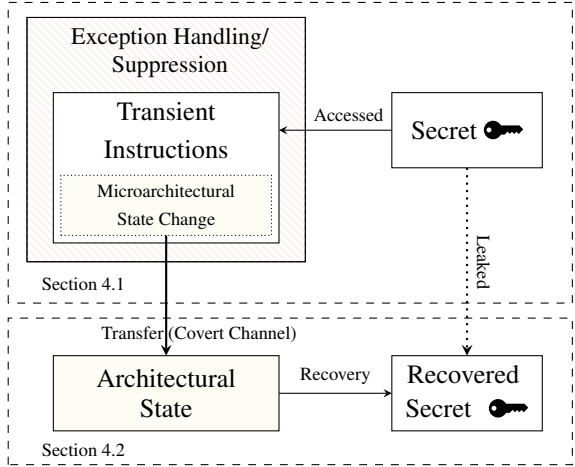

The full \AttackName attack consists of two building blocks, as illustrated in \Cref{fig:building-blocks}.
The first building block of \AttackName is to make the CPU execute one or more instructions that would never occur in the executed path.
In the toy example (\cf \Cref{sec:toy-example}), this is an access to an array, which would normally never be executed, as the previous instruction always raises an exception.
We call such an instruction, which is executed out of order, leaving measurable side effects, a \emph{\instructionname instruction}.
Furthermore, we call any sequence of instructions containing at least one \instructionname instruction a \instructionname instruction sequence.

In order to leverage \instructionname instructions for an attack, the \instructionname instruction sequence must utilize a secret value that an attacker wants to leak.
\Cref{sec:building-blocks:abandoned-instructions} describes building blocks to run a \instructionname instruction sequence with a dependency on a secret value.

The second building block of \AttackName is to transfer the microarchitectural side effect of the \instructionname instruction sequence to an architectural state to further process the leaked secret.
Thus, the second building described in \Cref{sec:building-blocks:covert-channel} describes building blocks to transfer a microarchitectural side effect to an architectural state using a covert channel.

\subsection{Executing \Instructionname Instructions}\label{sec:building-blocks:abandoned-instructions}
The first building block of \AttackName is the execution of \instructionname instructions.
\Instructionname instructions basically occur all the time, as the CPU continuously runs ahead of the current instruction to minimize the experienced latency and thus maximize the performance (\cf \cref{sec:background:spex}).
\Instructionname instructions introduce an exploitable side channel if their operation depends on a secret value.
We focus on addresses that are mapped within the attacker's process, \ie the user-accessible user space addresses as well as the user-inaccessible kernel space addresses.
Note that attacks targeting code that is executed within the context (\ie address space) of another process are possible~\cite{Kocher2017}, but out of scope in this work, since all physical memory (including the memory of other processes) can be read through the kernel address space anyway.

Accessing user-inaccessible pages, such as kernel pages, triggers an exception which generally terminates the application.
If the attacker targets a secret at a user-inaccessible address, the attacker has to cope with this exception.
We propose two approaches: With \textit{exception handling}, we catch the exception effectively occurring after executing the \instructionname instruction sequence, and with \textit{exception suppression}, we prevent the exception from occurring at all and instead redirect the control flow after executing the \instructionname instruction sequence.
We discuss these approaches in detail in the following.

\paragraph{Exception handling.}
A trivial approach is to fork the attacking application before accessing the invalid memory location that terminates the process, and only access the invalid memory location in the child process.
The CPU executes the \instructionname instruction sequence in the child process before crashing.
The parent process can then recover the secret by observing the microarchitectural state, \eg through a side-channel.

It is also possible to install a signal handler that will be executed if a certain exception occurs, in this specific case a segmentation fault.
This allows the attacker to issue the instruction sequence and prevent the application from crashing, reducing the overhead as no new process has to be created.

\paragraph{Exception suppression.}
A different approach to deal with exceptions is to prevent them from being raised in the first place.
Transactional memory allows to group memory accesses into one seemingly atomic operation, giving the option to roll-back to a previous state if an error occurs.
If an exception occurs within the transaction, the architectural state is reset, and the program execution continues without disruption.

Furthermore, speculative execution issues instructions that might not occur on the executed code path due to a branch misprediction.
Such instructions depending on a preceding conditional branch can be speculatively executed.
Thus, the invalid memory access is put within a speculative instruction sequence that is only executed if a prior branch condition evaluates to true.
By making sure that the condition never evaluates to true in the executed code path, we can suppress the occurring exception as the memory access is only executed speculatively.
This technique may require a sophisticated training of the branch predictor.
Kocher~\etal\cite{Kocher2017} pursue this approach in orthogonal work, since this construct can frequently be found in code of other processes.


\subsection{Building a Covert Channel}\label{sec:building-blocks:covert-channel}

The second building block of \AttackName is the transfer of the microarchitectural state, which was changed by the \instructionname instruction sequence, into an architectural state (\cf \Cref{fig:building-blocks}).
The \instructionname instruction sequence can be seen as the sending end of a microarchitectural covert channel.
The receiving end of the covert channel receives the microarchitectural state change and deduces the secret from the state.
Note that the receiver is not part of the \instructionname instruction sequence and can be a different thread or even a different process \eg the parent process in the fork-and-crash approach.

We leverage techniques from cache attacks, as the cache state is a microarchitectural state which can be reliably transferred into an architectural state using various techniques~\cite{Osvik2006,Yarom2014,Gruss2016Flush}.
Specifically, we use \FlushReload~\cite{Yarom2014}, as it allows to build a fast and low-noise covert channel.
Thus, depending on the secret value, the transient instruction sequence (\cf \Cref{sec:building-blocks:abandoned-instructions}) performs a regular memory access, \eg as it does in the toy example (\cf \Cref{sec:toy-example}).

After the transient instruction sequence accessed an accessible address, \ie this is the sender of the covert channel; the address is cached for subsequent accesses.
The receiver can then monitor whether the address has been loaded into the cache by measuring the access time to the address.
Thus, the sender can transmit a `1'-bit by accessing an address which is loaded into the monitored cache, and a `0'-bit by not accessing such an address.

Using multiple different cache lines, as in our toy example in \Cref{sec:toy-example}, allows to transmit multiple bits at once.
For every of the \SIx{256} different byte values, the sender accesses a different cache line.
By performing a \FlushReload attack on all of the \SIx{256} possible cache lines, the receiver can recover a full byte instead of just one bit.
However, since the \FlushReload attack takes much longer (typically several hundred cycles) than the transient instruction sequence, transmitting only a single bit at once is more efficient.
The attacker can simply do that by shifting and masking the secret value accordingly.

Note that the covert channel is not limited to microarchitectural states which rely on the cache.
Any microarchitectural state which can be influenced by an instruction (sequence) and is observable through a side channel can be used to build the sending end of a covert channel.
The sender could, for example, issue an instruction (sequence) which occupies a certain execution port such as the ALU to send a `1'-bit.
The receiver measures the latency when executing an instruction (sequence) on the same execution port.
A high latency implies that the sender sends a `1'-bit, whereas a low latency implies that sender sends a `0'-bit.
The advantage of the \FlushReload cache covert channel is the noise resistance and the high transmission rate~\cite{Gruss2016Flush}. Furthermore, the leakage can be observed from any CPU core~\cite{Yarom2014}, \ie rescheduling events do not significantly affect the covert channel.


\section{Meltdown}\label{sec:meltdown} 

In this section, present \AttackName, a powerful attack allowing to read arbitrary physical memory from an unprivileged user program, comprised of the building blocks presented in \Cref{sec:building-blocks}.
First, we discuss the attack setting to emphasize the wide applicability of this attack.
Second, we present an attack overview, showing how \AttackName can be mounted on both Windows and Linux on personal computers as well as in the cloud.
Finally, we discuss a concrete implementation of \AttackName allowing to dump kernel memory with up to \SI{503}{KB/\second}.

\paragraph{Attack setting.}
In our attack, we consider personal computers and virtual machines in the cloud.
In the attack scenario, the attacker has arbitrary unprivileged code execution on the attacked system, \ie the attacker can run any code with the privileges of a normal user.
However, the attacker has no physical access to the machine.
Further, we assume that the system is fully protected with state-of-the-art software-based defenses such as ASLR and KASLR as well as CPU features like SMAP, SMEP, NX, and PXN.
Most importantly, we assume a completely bug-free operating system, thus, no software vulnerability exists that can be exploited to gain kernel privileges or leak information.
The attacker targets secret user data, \eg passwords and private keys, or any other valuable information.

\subsection{Attack Description}\label{sec:meltdown:attack}

\AttackName combines the two building blocks discussed in \Cref{sec:building-blocks}.
First, an attacker makes the CPU execute a \instructionname instruction sequence which uses an inaccessible secret value stored somewhere in physical memory (\cf \Cref{sec:building-blocks:abandoned-instructions}).
The \instructionname instruction sequence acts as the transmitter of a covert channel (\cf \Cref{sec:building-blocks:covert-channel}), ultimately leaking the secret value to the attacker.

\AttackName consists of 3 steps:
\begin{compactenum}
 \item[\textbf{Step 1}] The content of an attacker-chosen memory location, which is inaccessible to the attacker, is loaded into a register.
 \item[\textbf{Step 2}] A \instructionname instruction accesses a cache line based on the secret content of the register.
 \item[\textbf{Step 3}] The attacker uses \FlushReload to determine the accessed cache line and hence the secret stored at the chosen memory location.
\end{compactenum}
By repeating these steps for different memory locations, the attacker can dump the kernel memory, including the entire physical memory.

\cref{lst:meltdown} shows the basic implementation of the \instructionname instruction sequence and the sending part of the covert channel, using x86 assembly instructions.
Note that this part of the attack could also be implemented entirely in higher level languages like C.
In the following, we will discuss each step of \AttackName and the corresponding code line in \Cref{lst:meltdown}.

\begin{lstlisting}[language={[x86masm]Assembler},style=customasm,float,caption={The core instruction sequence of \AttackName. An inaccessible kernel address is moved to a register, raising an exception. The subsequent instructions are already executed out of order before the exception is raised, leaking the content of the kernel address through the indirect memory access.}, label=lst:meltdown]
; rcx = kernel address
; rbx = probe array
retry:
mov al, byte [rcx]
shl rax, 0xc
jz retry
mov rbx, qword [rbx + rax]
\end{lstlisting}

\paragraph{Step 1: Reading the secret.}

To load data from the main memory into a register, the data in the main memory is referenced using a virtual address.
In parallel to translating a virtual address into a physical address, the CPU also checks the permission bits of the virtual address, \ie whether this virtual address is user accessible or only accessible by the kernel.
As already discussed in \Cref{sec:background:vaddr}, this hardware-based isolation through a permission bit is considered secure and recommended by the hardware vendors.
Hence, modern operating systems always map the entire kernel into the virtual address space of every user process.

As a consequence, all kernel addresses lead to a valid physical address when translating them, and the CPU can access the content of such addresses.
The only difference to accessing a user space address is that the CPU raises an exception as the current permission level does not allow to access such an address.
Hence, the user space cannot simply read the contents of such an address.
However, \AttackName exploits the out-of-order execution of modern CPUs, which still executes instructions in the small time window between the illegal memory access and the raising of the exception.

In line 4 of \Cref{lst:meltdown}, we load the byte value located at the target kernel address, stored in the \texttt{RCX} register, into the least significant byte of the \texttt{RAX} register represented by \texttt{AL}.
As explained in more detail in~\cref{sec:background:spex}, the \texttt{MOV} instruction is fetched by the core, decoded into \muops, allocated, and sent to the reorder buffer.
There, architectural registers (\eg \texttt{RAX} and \texttt{RCX} in \Cref{lst:meltdown}) are mapped to underlying physical registers enabling out-of-order execution.
Trying to utilize the pipeline as much as possible, subsequent instructions (lines 5-7) are already decoded and allocated as \muops as well.
The \muops are further sent to the reservation station holding the \muops while they wait to be executed by the corresponding execution unit.
The execution of a \muop can be delayed if execution units are already used to their corresponding capacity or operand values have not been calculated yet.

When the kernel address is loaded in line 4, it is likely that the CPU already issued the subsequent instructions as part of the out-or-order execution, and that their corresponding \muops wait in the reservation station for the content of the kernel address to arrive.
As soon as the fetched data is observed on the common data bus, the \muops can begin their execution.

When the \muops finish their execution, they retire in-order, and, thus, their results are committed to the architectural state.
During the retirement, any interrupts and exception that occurred during the execution of the instruction are handled.
Thus, if the \texttt{MOV} instruction that loads the kernel address is retired, the exception is registered and the pipeline is flushed to eliminate all results of subsequent instructions which were executed out of order.
However, there is a race condition between raising this exception and our attack step 2 which we describe below.

As reported by Gruss~\etal\cite{Gruss2016Prefetch}, prefetching kernel addresses sometimes succeeds.
We found that prefetching the kernel address can slightly improve the performance of the attack on some systems.

\paragraph{Step 2: Transmitting the secret.}

The instruction sequence from step 1 which is executed out of order has to be chosen in a way that it becomes a \instructionname instruction sequence.
If this \instructionname instruction sequence is executed before the \texttt{MOV} instruction is retired (\ie raises the exception), and the \instructionname instruction sequence performed computations based on the secret, it can be utilized to transmit the secret to the attacker.

As already discussed, we utilize cache attacks that allow to build fast and low-noise covert channel using the CPU's cache.
Thus, the \instructionname instruction sequence has to encode the secret into the microarchitectural cache state, similarly to the toy example in \Cref{sec:toy-example}.

We allocate a probe array in memory and ensure that no part of this array is cached.
To transmit the secret, the \instructionname instruction sequence contains an indirect memory access to an address which is calculated based on the secret (inaccessible) value.
In line 5 of \Cref{lst:meltdown} the secret value from step 1 is multiplied by the page size, \ie \SI{4}{KB}.
The multiplication of the secret ensures that accesses to the array have a large spatial distance to each other. 
This prevents the hardware prefetcher from loading adjacent memory locations into the cache as well.
Here, we read a single byte at once, hence our probe array is $256 \times 4096$ bytes, assuming \SI{4}{KB} pages.

Note that in the out-of-order execution we have a noise-bias towards register value `0'.
We discuss the reasons for this in \Cref{sec:attack:zero}.
However, for this reason, we introduce a retry-logic into the \instructionname instruction sequence.
In case we read a `0', we try to read the secret again (step 1).
In line 7, the multiplied secret is added to the base address of the probe array, forming the target address of the covert channel.
This address is read to cache the corresponding cache line.
Consequently, our \instructionname instruction sequence affects the cache state based on the secret value that was read in step 1.

Since the \instructionname instruction sequence in step 2 races against raising the exception, reducing the runtime of step 2 can significantly improve the performance of the attack.
For instance, taking care that the address translation for the probe array is cached in the TLB increases the attack performance on some systems.

\paragraph{Step 3: Receiving the secret.}
In step 3, the attacker recovers the secret value (step 1) by leveraging a microarchitectural side-channel attack (\ie the receiving end of a microarchitectural covert channel) that transfers the cache state (step 2) back into an architectural state.
As discussed in \Cref{sec:building-blocks:covert-channel}, \AttackName relies on \FlushReload to transfer the cache state into an architectural state.

When the \instructionname instruction sequence of step 2 is executed, exactly one cache line of the probe array is cached.
The position of the cached cache line within the probe array depends only on the secret which is read in step 1.
Thus, the attacker iterates over all 256 pages of the probe array and measures the access time for every first cache line (\ie offset) on the page.
The number of the page containing the cached cache line corresponds directly to the secret value.

\paragraph{Dumping the entire physical memory.}
By repeating all 3 steps of \AttackName, the attacker can dump the entire memory by iterating over all different addresses.
However, as the memory access to the kernel address raises an exception that terminates the program, we use one of the methods described in \Cref{sec:building-blocks:abandoned-instructions} to handle or suppress the exception.

As all major operating systems also typically map the entire physical memory into the kernel address space (\cf \cref{sec:background:vaddr}) in every user process, \AttackName is not only limited to reading kernel memory but it is capable of reading the entire physical memory of the target machine.

\subsection{Optimizations and Limitations}

\paragraph{The case of 0.}\label{sec:attack:zero}
If the exception is triggered while trying to read from an inaccessible kernel address, the register where the data should be stored, appears to be zeroed out.
This is reasonable because if the exception is unhandled, the user space application is terminated, and the value from the inaccessible kernel address could be observed in the register contents stored in the core dump of the crashed process.
The direct solution to fix this problem is to zero out the corresponding registers.
If the zeroing out of the register is faster than the execution of the subsequent instruction (line 5 in \Cref{lst:meltdown}), the attacker may read a false value in the third step.
To prevent the \instructionname instruction sequence from continuing with a wrong value, \ie `0', \AttackName retries reading the address until it encounters a value different from `0' (line 6).
As the \instructionname instruction sequence terminates after the exception is raised, there is no cache access if the secret value is 0.
Thus, \AttackName assumes that the secret value is indeed `0' if there is no cache hit at all.

The loop is terminated by either the read value not being `0' or by the raised exception of the invalid memory access.
Note that this loop does not slow down the attack measurably, since, in either case, the processor runs ahead of the illegal memory access, regardless of whether ahead is a loop or ahead is a linear control flow.
In either case, the time until the control flow returned from exception handling or exception suppression remains the same with and without this loop.
Thus, capturing read `0's beforehand and recovering early from a lost race condition vastly increases the reading speed.

\paragraph{Single-bit transmission}
In the attack description in \cref{sec:meltdown:attack}, the attacker transmitted 8 bits through the covert channel at once and performed $2^8=256$ \FlushReload measurements to recover the secret.
However, there is a clear trade-off between running more \instructionname instruction sequences and performing more \FlushReload measurements.
The attacker could transmit an arbitrary number of bits in a single transmission through the covert channel, by either reading more bits using a \texttt{MOV} instruction for a larger data value.
Furthermore, the attacker could mask bits using additional instructions in the \instructionname instruction sequence.
We found the number of additional instructions in the \instructionname instruction sequence to have a negligible influence on the performance of the attack.

The performance bottleneck in the generic attack description above is indeed, the time spent on \FlushReload measurements.
In fact, with this implementation, almost the entire time will be spent on \FlushReload measurements.
By transmitting only a single bit, we can omit all but one \FlushReload measurement, \ie the measurement on cache line 1.
If the transmitted bit was a `1', then we observe a cache hit on cache line 1.
Otherwise, we observe no cache hit on cache line 1.

Transmitting only a single bit at once also has drawbacks.
As described above, our side channel has a bias towards a secret value of `0'.
If we read and transmit multiple bits at once, the likelihood that all bits are `0' may quite small for actual user data.
The likelihood that a single bit is `0' is typically close to \SI{50}{\percent}.
Hence, the number of bits read and transmitted at once is a trade-off between some implicit error-reduction and the overall transmission rate of the covert channel.

However, since the error rates are quite small in either case, our evaluation (\cf \cref{sec:evaluation}) is based on the single-bit transmission mechanics.

\paragraph{Exception Suppression using Intel TSX.}

In~\cref{sec:building-blocks:abandoned-instructions}, we discussed the option to prevent that an exception is raised due an invalid memory access in the first place.
Using Intel TSX, a hardware transactional memory implementation, we can completely suppress the exception~\cite{Jang2016}.

With Intel TSX, multiple instructions can be grouped to a transaction, which appears to be an atomic operation, \ie either all or no instruction is executed.
If one instruction within the transaction fails, already executed instructions are reverted, but no exception is raised.

If we wrap the code from \cref{lst:meltdown} with such a TSX instruction, any exception is suppressed.
However, the microarchitectural effects are still visible, \ie the cache state is persistently manipulated from within the hardware transaction~\cite{Gruss2017TSX}.
This results in a higher channel capacity, as suppressing the exception is significantly faster than trapping into the kernel for handling the exception, and continuing afterwards.

\paragraph{Dealing with KASLR.}\label{sec:attack:kaslr}
In 2013, kernel address space layout randomization (KASLR) had been introduced to the Linux kernel (starting from version 3.14~\cite{Edge2013Kaslr}) allowing to randomize the location of the kernel code at boot time.
However, only as recently as May 2017, KASLR had been enabled by default in version 4.12~\cite{Molnar2017Kaslr}.
With KASLR also the direct-physical map is randomized and, thus, not fixed at a certain address such that the attacker is required to obtain the randomized offset before mounting the \AttackName attack.
However, the randomization is limited to \SIx{40} bit.

Thus, if we assume a setup of the target machine with \SI{8}{\giga B} of RAM, it is sufficient to test the address space for addresses in \SI{8}{\giga B} steps.
This allows to cover the search space of \SIx{40} bit with only \SIx{128} tests in the worst case.
If the attacker can successfully obtain a value from a tested address, the attacker can proceed dumping the entire memory from that location.
This allows to mount \AttackName on a system despite being protected by KASLR within seconds.


\section{Evaluation}\label{sec:evaluation} 

In this section, we evaluate \AttackName and the performance of our proof-of-concept implementation~\footnote{\url{https://github.com/IAIK/meltdown}}.
\Cref{sec:eval:results} discusses the information which \AttackName can leak, and \Cref{sec:eval:performance} evaluates the performance of \AttackName, including countermeasures.
Finally, we discuss limitations for AMD and ARM in \Cref{sec:eval:negative}.

\begin{table}
	\centering
	\caption{Experimental setups.}\label{tab:setup}
	\begin{tabular}{clr}
		\toprule
		Environment & CPU model & Cores \\
		\midrule
		Lab & Celeron G540 & 2 \\
		Lab & Core i5-3230M & 2 \\
		Lab & Core i5-3320M & 2 \\
		Lab & Core i7-4790 & 4 \\
		Lab & Core i5-6200U & 2 \\
		Lab & Core i7-6600U & 2  \\
		Lab & Core i7-6700K & 4 \\
		Cloud & Xeon E5-2676 v3 & 12 \\
		Cloud & Xeon E5-2650 v4 & 12 \\
		\bottomrule
	\end{tabular}
\end{table}

\Cref{tab:setup} shows a list of configurations on which we successfully reproduced \AttackName.
For the evaluation of \AttackName, we used both laptops as well as desktop PCs with Intel Core CPUs.
For the cloud setup, we tested \AttackName in virtual machines running on Intel Xeon CPUs hosted in the Amazon Elastic Compute Cloud as well as on DigitalOcean.
Note that for ethical reasons we did not use \AttackName on addresses referring to physical memory of other tenants.

\subsection{Information Leakage and Environments}\label{sec:eval:results}

We evaluated \AttackName on both Linux (\cf \Cref{sec:eval:linux}) and Windows 10 (\cf \Cref{sec:eval:windows}).
On both operating systems, \AttackName can successfully leak kernel memory.
Furthermore, we also evaluated the effect of the KAISER patches on \AttackName on Linux, to show that KAISER prevents the leakage of kernel memory (\cf \cref{sec:eval:linux_kaiser}).
Finally, we discuss the information leakage when running inside containers such as Docker (\cf \cref{sec:eval:docker}).

\subsubsection{Linux}\label{sec:eval:linux}

We successfully evaluated \AttackName on multiple versions of the Linux kernel, from 2.6.32 to 4.13.0.
On all these versions of the Linux kernel, the kernel address space is also mapped into the user address space.
Thus, all kernel addresses are also mapped into the address space of user space applications, but any access is prevented due to the permission settings for these addresses.
As \AttackName bypasses these permission settings, an attacker can leak the complete kernel memory if the virtual address of the kernel base is known.
Since all major operating systems also map the entire physical memory into the kernel address space (\cf \cref{sec:background:vaddr}), all physical memory can also be read.

Before kernel 4.12, kernel address space layout randomization (KASLR) was not active by default~\cite{Phoronix2017KASLR}.
If KASLR is active, \AttackName can still be used to find the kernel by searching through the address space (\cf \Cref{sec:attack:kaslr}).
An attacker can also simply de-randomize the direct-physical map by iterating through the virtual address space.
Without KASLR, the direct-physical map starts at address \texttt{0xffff 8800 0000 0000} and linearly maps the entire physical memory.
On such systems, an attacker can use \AttackName to dump the entire physical memory, simply by reading from virtual addresses starting at \texttt{0xffff 8800 0000 0000}.

On newer systems, where KASLR is active by default, the randomization of the direct-physical map is limited to \SIx{40} bit.
It is even further limited due to the linearity of the mapping.
Assuming that the target system has at least \SI{8}{\giga B} of physical memory, the attacker can test addresses in steps of \SI{8}{\giga B}, resulting in a maximum of \SIx{128} memory locations to test.
Starting from one discovered location, the attacker can again dump the entire physical memory.

Hence, for the evaluation, we can assume that the randomization is either disabled, or the offset was already retrieved in a pre-computation step.

\subsubsection{Linux with KAISER Patch}\label{sec:eval:linux_kaiser}

The KAISER patch by Gruss~\etal\cite{Gruss2017Kaslr} implements a stronger isolation between kernel and user space.
KAISER does not map any kernel memory in the user space, except for some parts required by the x86 architecture (\eg interrupt handlers).
Thus, there is no valid mapping to either kernel memory or physical memory (via the direct-physical map) in the user space, and such addresses can therefore not be resolved.
Consequently, \AttackName cannot leak any kernel or physical memory except for the few memory locations which have to be mapped in user space.

We verified that KAISER indeed prevents \AttackName, and there is no leakage of any kernel or physical memory.

Furthermore, if KASLR is active, and the few remaining memory locations are randomized, finding these memory locations is not trivial due to their small size of several kilobytes.
\Cref{sec:kaiser} discusses the implications of these mapped memory locations from a security perspective.

\subsubsection{Microsoft Windows}\label{sec:eval:windows}

We successfully evaluated \AttackName on an up-to-date Microsoft Windows 10 operating system.
In line with the results on Linux (\cf \Cref{sec:eval:linux}), \AttackName also can leak arbitrary kernel memory on Windows.
This is not surprising, since \AttackName does not exploit any software issues, but is caused by a hardware issue.

In contrast to Linux, Windows does not have the concept of an identity mapping, which linearly maps the physical memory into the virtual address space.
Instead, a large fraction of the physical memory is mapped in the paged pools, non-paged pools, and the system cache.
Furthermore, Windows maps the kernel into the address space of every application too.
Thus, \AttackName can read kernel memory which is mapped in the kernel address space, \ie any part of the kernel which is not swapped out, and any page mapped in the paged and non-paged pool, and the system cache.

Note that there likely are physical pages which are mapped in one process but not in the (kernel) address space of another process, \ie physical pages which cannot be attacked using \AttackName.
However, most of the physical memory will still be accessible through \AttackName.

We were successfully able to read the binary of the Windows kernel using \AttackName.
To verify that the leaked data is actual kernel memory, we first used the Windows kernel debugger to obtain kernel addresses containing actual data.
After leaking the data, we again used the Windows kernel debugger to compare the leaked data with the actual memory content, confirming that \AttackName can successfully leak kernel memory.


\subsubsection{Containers}\label{sec:eval:docker}

We evaluated \AttackName running in containers sharing a kernel, including Docker, LXC, and OpenVZ, and found that the attack can be mounted without any restrictions.
Running \AttackName inside a container allows to leak information not only from the underlying kernel, but also from all other containers running on the same physical host.

The commonality of most container solutions is that every container uses the same kernel, \ie the kernel is shared among all containers.
Thus, every container has a valid mapping of the entire physical memory through the direct-physical map of the shared kernel.
Furthermore, \AttackName cannot be blocked in containers, as it uses only memory accesses.
Especially with Intel TSX, only unprivileged instructions are executed without even trapping into the kernel.

Thus, the isolation of containers sharing a kernel can be fully broken using \AttackName.
This is especially critical for cheaper hosting providers where users are not separated through fully virtualized machines, but only through containers.
We verified that our attack works in such a setup, by successfully leaking memory contents from a container of a different user under our control.

\subsection{\AttackName Performance}\label{sec:eval:performance}

To evaluate the performance of \AttackName, we leaked known values from kernel memory.
This allows us to not only determine how fast an attacker can leak memory, but also the error rate, \ie how many byte errors to expect.
We achieved average reading rates of up to \SI{503}{KB/\second} with an error rate as low as \SI{0.02}{\percent} when using exception suppression.
For the performance evaluation, we focused on the Intel Core i7-6700K as it supports Intel TSX, to get a fair performance comparison between exception handling and exception suppression.

For all tests, we use \FlushReload as a covert channel to leak the memory as described in \Cref{sec:meltdown}.
We evaluated the performance of both exception handling and exception suppression (\cf \Cref{sec:building-blocks:abandoned-instructions}).
For exception handling, we used signal handlers, and if the CPU supported it, we also used exception suppression using Intel TSX.
An extensive evaluation of exception suppression using conditional branches was done by Kocher~\etal\cite{Kocher2017} and is thus omitted in this paper for the sake of brevity.

\subsubsection{Exception Handling}

Exception handling is the more universal implementation, as it does not depend on any CPU extension and can thus be used without any restrictions.
The only requirement for exception handling is operating system support to catch segmentation faults and continue operation afterwards.
This is the case for all modern operating systems, even though the specific implementation differs between the operating systems.
On Linux, we used signals, whereas, on Windows, we relied on the Structured Exception Handler.

With exception handling, we achieved average reading speeds of \SI{123}{KB/\second} when leaking \SI{12}{\mega B} of kernel memory.
Out of the \SI{12}{\mega B} kernel data, only \SI{0.03}{\percent} were read incorrectly.
Thus, with an error rate of \SI{0.03}{\percent}, the channel capacity is \SI{122}{KB/\second}.

\subsubsection{Exception Suppression}

Exception suppression can either be achieved using conditional branches or using Intel TSX.
Conditional branches are covered in detail in Kocher~\etal\cite{Kocher2017}, hence we only evaluate Intel TSX for exception suppression.
In contrast to exception handling, Intel TSX does not require operating system support, as it is an instruction-set extension.
However, Intel TSX is a rather new extension and is thus only available on recent Intel CPUs, \ie since the Broadwell microarchitecture.

Again, we leaked \SI{12}{\mega B} of kernel memory to measure the performance.
With exception suppression, we achieved average reading speeds of \SI{503}{KB/\second}.
Moreover, the error rate of \SI{0.02}{\percent} with exception suppression is even lower than with exception handling.
Thus, the channel capacity we achieve with exception suppression is \SI{502}{KB/\second}.

\subsection{\AttackName in Practice}\label{sec:eval:examples}

\begin{figure}[t]
 \centering
 \tikzsetnextfilename{ff-pwds}
 \resizebox{\hsize}{!}{%
 \includegraphics{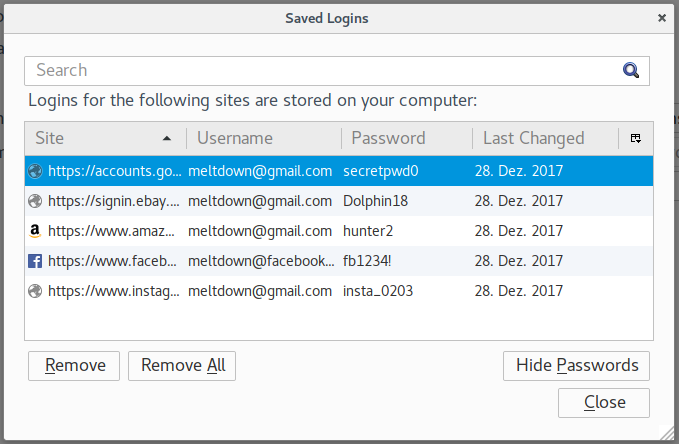}
 }%
 \caption{Firefox 56 password manager showing the stored passwords that are leaked using \AttackName in \Cref{lst:memory-dump-firefox}.}
 \label{fig:ff-pwds}
\end{figure}

\begin{lstlisting}[language=C,style=customc,float,caption={Memory dump showing HTTP Headers on Ubuntu 16.10 on a Intel Core i7-6700K}, float, label=lst:memory-dump,numbers=none,basicstyle=\tiny\ttfamily,xleftmargin=3.5pt,xrightmargin=-3.5pt,framexleftmargin=0pt,framexrightmargin=-7.5pt]
79cbb30: 616f 61 4e 6b 32 38 46 31  34 67 65 68 61 7a 34 |aoaNk28F14gehaz4|
79cbb40: 5a74 4d 79 78 68 76 41 57  69 69 63 77 59 62 61 |ZtMyxhvAWiicwYba|
79cbb50: 356a 4c 76 4d 70 4b 56 56  32 4b 6a 37 4b 5a 4e |5jLvMpKVV2Kj7KZN|
79cbb60: 6655 6c 6e 72 38 64 74 35  54 62 43 63 7a 6f 44 |fUlnr8dt5TbCczoD|
79cbb70: 494e 46 71 58 6d 4a 69 34  58 50 39 62 43 53 47 |INFqXmJi4XP9bCSG|
79cbb80: 6c4c 48 32 5a 78 66 56 44  73 4b 57 39 34 68 6d |lLH2ZxfVDsKW94hm|
79cbb90: 3364 2f 41 4d 41 45 44 41  41 41 41 41 51 45 42 |3d/AMAEDAAAAAQEB|
79cbba0: 4141 41 41 41 41 3d 3d XX  XX XX XX XX XX XX XX |AAAAAA==........|
79cbbb0: XXXX XX XX XX XX XX XX XX  XX XX XX XX XX XX XX |................|
79cbbc0: XXXX XX 65 2d 68 65 61 64  XX XX XX XX XX XX XX |...e-head.......|
79cbbd0: XXXX XX XX XX XX XX XX XX  XX XX XX XX XX XX XX |................|
79cbbe0: XXXX XX XX XX XX XX XX XX  XX XX XX XX XX XX XX |................|
79cbbf0: XXXX XX XX XX XX XX XX XX  XX XX XX XX XX XX XX |................|
79cbc00: XXXX XX XX XX XX XX XX XX  XX XX XX XX XX XX XX |................|
79cbc10: XXXX XX XX XX XX XX XX XX  XX XX XX XX XX XX XX |................|
79cbc20: XXXX XX XX XX XX XX XX XX  XX XX XX XX XX XX XX |................|
79cbc30: XXXX XX XX XX XX XX XX XX  XX XX XX XX XX XX XX |................|
79cbc40: XXXX XX XX XX XX XX XX XX  XX XX XX XX XX XX XX |................|
79cbc50: XXXX XX XX 0d 0a XX 6f 72  69 67 69 6e 61 6c 2d |.......original-|
79cbc60: 7265 73 70 6f 6e 73 65 2d  68 65 61 64 65 72 73 |response-headers|
79cbc70: XX44 61 74 65 3a 20 53 61  74 2c 20 30 39 20 44 |.Date: Sat, 09 D|
79cbc80: 6563 20 32 30 31 37 20 32  32 3a 32 39 3a 32 35 |ec 2017 22:29:25|
79cbc90: 2047 4d 54 0d 0a 43 6f 6e  74 65 6e 74 2d 4c 65 | GMT..Content-Le|
79cbca0: 6e67 74 68 3a 20 31 0d 0a  43 6f 6e 74 65 6e 74 |ngth: 1..Content|
79cbcb0: 2d54 79 70 65 3a 20 74 65  78 74 2f 68 74 6d 6c |-Type: text/html|
79cbcc0: 3b20 63 68 61 72 73 65 74  3d 75 74 66 2d 38 0d |; charset=utf-8.|
79cbcd0: 0a53 65 72 76 65 72 3a 20  54 77 69 73 74 65 64 |.Server: Twisted|
79cbce0: 5765 62 2f 31 36 2e 33 2e  30 0d 0a XX 75 6e 63 |Web/16.3.0...unc|
79cbcf0: 6f6d 70 72 65 73 73 65 64  2d 6c 65 6e XX XX XX |ompressed-len...|
\end{lstlisting}

\begin{lstlisting}[language=C,style=customc,float,caption={Memory dump of Firefox 56 on Ubuntu 16.10 on a Intel Core i7-6700K disclosing saved passwords (\cf \Cref{fig:ff-pwds}).}, float, label=lst:memory-dump-firefox,numbers=none,basicstyle=\tiny\ttfamily,xleftmargin=3.5pt,xrightmargin=-3.5pt,framexleftmargin=0pt,framexrightmargin=-7.5pt]
f94b7690: e5 e5 e5 e5 e5 e5 e5 e5 e5  e5 e5 e5 e5 e5 e5 e5 |................|
f94b76a0: e5 e5 e5 e5 e5 e5 e5 e5 e5  e5 e5 e5 e5 e5 e5 e5 |................|
f94b76b0: 70 52 b8 6b 96 7f XX XX XX  XX XX XX XX XX XX XX |pR.k............|
f94b76c0: 09 XX XX XX XX XX XX XX XX  XX XX XX XX XX XX XX |................|
f94b76d0: XX XX XX XX XX XX XX XX XX  XX XX XX XX XX XX XX |................|
f94b76e0: XX XX XX XX XX XX XX XX XX  XX XX XX XX XX XX 81 |................|
f94b76f0: 12 XX e0 81 19 XX e0 81 44  6f 6c 70 68 69 6e 31 |........Dolphin1|
f94b7700: 38 e5 e5 e5 e5 e5 e5 e5 e5  e5 e5 e5 e5 e5 e5 e5 |8...............|
f94b7710: 70 52 b8 6b 96 7f XX XX XX  XX XX XX XX XX XX XX |pR.k............|
f94b7720: XX XX XX XX XX XX XX XX XX  XX XX XX XX XX XX XX |................|
f94b7730: XX XX XX XX 4a XX XX XX XX  XX XX XX XX XX XX XX |....J...........|
f94b7740: XX XX XX XX XX XX XX XX XX  XX XX XX XX XX XX XX |................|
f94b7750: XX XX XX XX XX XX XX XX XX  XX e0 81 69 6e 73 74 |............inst|
f94b7760: 61 5f 30 32 30 33 e5 e5 e5  e5 e5 e5 e5 e5 e5 e5 |a_0203..........|
f94b7770: 70 52 18 7d 28 7f XX XX XX  XX XX XX XX XX XX XX |pR.}(...........|
f94b7780: XX XX XX XX XX XX XX XX XX  XX XX XX XX XX XX XX |................|
f94b7790: XX XX XX XX 54 XX XX XX XX  XX XX XX XX XX XX XX |....T...........|
f94b77a0: XX XX XX XX XX XX XX XX XX  XX XX XX XX XX XX XX |................|
f94b77b0: XX XX XX XX XX XX XX XX XX  XX XX XX 73 65 63 72 |............secr|
f94b77c0: 65 74 70 77 64 30 e5 e5 e5  e5 e5 e5 e5 e5 e5 e5 |etpwd0..........|
f94b77d0: 30 b4 18 7d 28 7f XX XX XX  XX XX XX XX XX XX XX |0..}(...........|
f94b77e0: XX XX XX XX XX XX XX XX XX  XX XX XX XX XX XX XX |................|
f94b77f0: XX XX XX XX XX XX XX XX XX  XX XX XX XX XX XX XX |................|
f94b7800: e5 e5 e5 e5 e5 e5 e5 e5 e5  e5 e5 e5 e5 e5 e5 e5 |................|
f94b7810: 68 74 74 70 73 3a 2f 2f 61  64 64 6f 6e 73 2e 63 |https://addons.c|
f94b7820: 64 6e 2e 6d 6f 7a 69 6c 6c  61 2e 6e 65 74 2f 75 |dn.mozilla.net/u|
f94b7830: 73 65 72 2d 6d 65 64 69 61  2f 61 64 64 6f 6e 5f |ser-media/addon_|
f94b7840: 69 63 6f 6e 73 2f 33 35 34  2f 33 35 34 33 39 39 |icons/354/354399|
f94b7850: 2d 36 34 2e 70 6e 67 3f 6d  6f 64 69 66 69 65 64 |-64.png?modified|
f94b7860: 3d 31 34 35 32 32 34 34 38  31 35 XX XX XX XX XX |=1452244815.....|
\end{lstlisting}

\cref{lst:memory-dump} shows a memory dump using \AttackName on an Intel Core i7-6700K running Ubuntu 16.10 with the Linux kernel 4.8.0.
In this example, we can identify HTTP headers of a request to a web server running on the machine.
The XX cases represent bytes where the side channel did not yield any results, \ie no \FlushReload hit.
Additional repetitions of the attack may still be able to read these bytes.

\Cref{lst:memory-dump-firefox} shows a memory dump of Firefox 56 using \AttackName on the same machine.
We can clearly identify some of the passwords that are stored in the internal password manager shown in \Cref{fig:ff-pwds}, \ie \texttt{Dolphin18}, \texttt{insta\_0203}, and \texttt{secretpwd0}.
The attack also recovered a URL which appears to be related to a Firefox addon.

\subsection{Limitations on ARM and AMD}\label{sec:eval:negative}

We also tried to reproduce the \AttackName bug on several ARM and AMD CPUs.
However, we did not manage to successfully leak kernel memory with the attack described in~\cref{sec:meltdown}, neither on ARM nor on AMD.
The reasons for this can be manifold.
First of all, our implementation might simply be too slow and a more optimized version might succeed.
For instance, a more shallow out-of-order execution pipeline could tip the race condition towards against the data leakage.
Similarly, if the processor lacks certain features, \eg no re-order buffer, our current implementation might not be able to leak data.
However, for both ARM and AMD, the toy example as described in \cref{sec:toy-example} works reliably, indicating that out-of-order execution generally occurs and instructions past illegal memory accesses are also performed.


\section{Countermeasures}
\label{sec:countermeasures}

In this section, we discuss countermeasures against the \AttackName attack.
At first, as the issue is rooted in the hardware itself, we want to discuss possible microcode updates and general changes in the hardware design.
Second, we want to discuss the \DefenseName countermeasure that has been developed to mitigate side-channel attacks against KASLR which inadvertently also protects against \AttackName.

\subsection{Hardware}\label{sec:countermeasure_hardware}
\AttackName bypasses the hardware-enforced isolation of security domains.
There is no software vulnerability involved in \AttackName.
Hence any software patch (\eg KAISER~\cite{Gruss2017Kaslr}) will leave small amounts of memory exposed (\cf \cref{sec:kaiser}).
There is no documentation whether such a fix requires the development of completely new hardware, or can be fixed using a microcode update.

As \AttackName exploits out-of-order execution, a trivial countermeasure would be to completely disable out-of-order execution.
However, the performance impacts would be devastating, as the parallelism of modern CPUs could not be leveraged anymore.
Thus, this is not a viable solution.

\AttackName is some form of race condition between the fetch of a memory address and the corresponding permission check for this address.
Serializing the permission check and the register fetch can prevent \AttackName, as the memory address is never fetched if the permission check fails.
However, this involves a significant overhead to every memory fetch, as the memory fetch has to stall until the permission check is completed.

A more realistic solution would be to introduce a hard split of user space and kernel space.
This could be enabled optionally by modern kernels using a new hard-split bit in a CPU control register, \eg CR4.
If the hard-split bit is set, the kernel has to reside in the upper half of the address space, and the user space has to reside in the lower half of the address space.
With this hard split, a memory fetch can immediately identify whether such a fetch of the destination would violate a security boundary, as the privilege level can be directly derived from the virtual address without any further lookups.
We expect the performance impacts of such a solution to be minimal.
Furthermore, the backwards compatibility is ensured, since the hard-split bit is not set by default and the kernel only sets it if it supports the hard-split feature.

Note that these countermeasures only prevent \AttackName, and not the class of Spectre attacks described by Kocher~\etal\cite{Kocher2017}.
Likewise, several countermeasures presented by Kocher~\etal\cite{Kocher2017} have no effect on \AttackName.
We stress that it is important to deploy countermeasures against both attacks.

\subsection{\DefenseName}\label{sec:kaiser}
As hardware is not as easy to patch, there is a need for software workarounds until new hardware can be deployed.
Gruss~\etal\cite{Gruss2017Kaslr} proposed \DefenseName, a kernel modification to not have the kernel mapped in the user space.
This modification was intended to prevent side-channel attacks breaking KASLR~\cite{Hund2013,Gruss2016Prefetch,Jang2016}.
However, it also prevents \AttackName, as it ensures that there is no valid mapping to kernel space or physical memory available in user space.
\DefenseName will be available in the upcoming releases of the Linux kernel under the name kernel page-table isolation (KPTI)~\cite{LWN_kpti}.
The patch will also be backported to older Linux kernel versions.
A similar patch was also introduced in Microsoft Windows 10 Build 17035~\cite{Ionescu2017Twitter}.
Also, Mac OS X and iOS have similar features~\cite{Levin2012}.

Although \DefenseName provides basic protection against \AttackName, it still has some limitations.
Due to the design of the x86 architecture, several privileged memory locations are required to be mapped in user space~\cite{Gruss2017Kaslr}.
This leaves a residual attack surface for \AttackName, \ie these memory locations can still be read from user space.
Even though these memory locations do not contain any secrets, such as credentials, they might still contain pointers.
Leaking one pointer can be enough to again break KASLR, as the randomization can be calculated from the pointer value.

Still, \DefenseName is the best short-time solution currently available and should therefore be deployed on all systems immediately.
Even with \AttackName, \DefenseName can avoid having any kernel pointers on memory locations that are mapped in the user space which would leak information about the randomized offsets.
This would require trampoline locations for every kernel pointer, \ie the interrupt handler would not call into kernel code directly, but through a trampoline function.
The trampoline function must only be mapped in the kernel.
It must be randomized with a different offset than the remaining kernel.
Consequently, an attacker can only leak pointers to the trampoline code, but not the randomized offsets of the remaining kernel.
Such trampoline code is required for every kernel memory that still has to be mapped in user space and contains kernel addresses.
This approach is a trade-off between performance and security which has to be assessed in future work.



\section{Discussion}\label{sec:discussion} 
\AttackName fundamentally changes our perspective on the security of hardware optimizations that manipulate the state of microarchitectural elements.
The fact that hardware optimizations can change the state of microarchitectural elements, and thereby imperil secure software implementations, is known since more than 20 years~\cite{Kocher1996}.
Both industry and the scientific community so far accepted this as a necessary evil for efficient computing.
Today it is considered a bug when a cryptographic algorithm is not protected against the microarchitectural leakage introduced by the hardware optimizations.
\AttackName changes the situation entirely.
\AttackName shifts the granularity from a comparably low spatial and temporal granularity, \eg 64-bytes every few hundred cycles for cache attacks, to an arbitrary granularity, allowing an attacker to read every single bit.
This is nothing any (cryptographic) algorithm can protect itself against.
\DefenseName is a short-term software fix, but the problem we uncovered is much more significant.

We expect several more performance optimizations in modern CPUs which affect the microarchitectural state in some way, not even necessarily through the cache.
Thus, hardware which is designed to provide certain security guarantees, \eg CPUs running untrusted code, require a redesign to avoid \AttackName- and Spectre-like attacks.
\AttackName also shows that even error-free software, which is explicitly written to thwart side-channel attacks, is not secure if the design of the underlying hardware is not taken into account.

With the integration of \DefenseName into all major operating systems, an important step has already been done to prevent \AttackName.
\DefenseName is also the first step of a paradigm change in operating systems.
Instead of always mapping everything into the address space, mapping only the minimally required memory locations appears to be a first step in reducing the attack surface.
However, it might not be enough, and an even stronger isolation may be required.
In this case, we can trade flexibility for performance and security, by \eg forcing a certain virtual memory layout for every operating system.
As most modern operating system already use basically the same memory layout, this might be a promising approach.

\AttackName also heavily affects cloud providers, especially if the guests are not fully virtualized.
For performance reasons, many hosting or cloud providers do not have an abstraction layer for virtual memory.
In such environments, which typically use containers, such as Docker or OpenVZ, the kernel is shared among all guests.
Thus, the isolation between guests can simply be circumvented with \AttackName, fully exposing the data of all other guests on the same host.
For these providers, changing their infrastructure to full virtualization or using software workarounds such as \DefenseName would both increase the costs significantly.

Even if \AttackName is fixed, Spectre~\cite{Kocher2017} will remain an issue.
Spectre~\cite{Kocher2017} and \AttackName need different defenses.
Specifically mitigating only one of them will leave the security of the entire system at risk.
We expect that \AttackName and Spectre open a new field of research to investigate in what extent performance optimizations change the microarchitectural state, how this state can be translated into an architectural state, and how such attacks can be prevented.


\section{Conclusion}\label{sec:conclusion} 
In this paper, we presented \AttackName, a novel software-based side-channel attack exploiting out-of-order execution on Intel CPUs to read arbitrary kernel- and physical-memory  locations from an unprivileged user space program.
Without requiring any software vulnerability and independent of the operating system, \AttackName~enables an adversary to read sensitive data of other processes or virtual machines in the cloud with up to \SI{503}{KB/\second}, affecting millions of devices.
We showed that the countermeasure \DefenseName~\cite{Gruss2017Kaslr}, originally proposed to protect from side-channel attacks against KASLR, inadvertently impedes \AttackName as well.
We stress that \DefenseName~needs to be deployed on every operating system as a short-term workaround, until \AttackName is fixed in hardware, to prevent large-scale exploitation of \AttackName.


\section*{Acknowledgment}
We would like to thank Anders Fogh for fruitful discussions at BlackHat USA 2016 and BlackHat Europe 2016, which ultimately led to the discovery of \AttackName.
Fogh~\cite{Fogh2017Speculative} already suspected that it might be possible to abuse speculative execution in order to read kernel memory in user mode but his experiments were not successful.
We would also like to thank Jann Horn for comments on an early draft.
Jann disclosed the issue to Intel in June.
The subsequent activity around the KAISER patch was the reason we started investigating this issue.
Furthermore, we would like Intel, ARM, Qualcomm, and Microsoft for feedback on an early draft.

We would also like to thank Intel for awarding us with a bug bounty for the responsible disclosure process, and their professional handling of this issue through communicating a clear timeline and connecting all involved researchers.
Furthermore, we would also thank ARM for their fast response upon disclosing the issue.

This work was supported in part by the European Research Council (ERC) under the European Union’s Horizon 2020 research and innovation programme (grant agreement No 681402).


{\footnotesize \bibliographystyle{acm}
\bibliography{main}}



\end{document}

%% file: images/core-skylake.tikz
\begin{tikzpicture}

\tikzstyle{execution_engine}+=[fill=green!30];
\tikzstyle{reorder_buffer}+=[fill=green!80];
\tikzstyle{scheduler}+=[fill=blue!30];
\tikzstyle{execution_units}+=[fill=yellow!50];
\tikzstyle{execution_unit}+=[fill=orange!50];

\tikzstyle{memory_subsystem}+=[fill=red!20];
\tikzstyle{tlb}+=[fill=yellow!20];
\tikzstyle{cache}+=[fill=red!40];

\tikzstyle{frontend}+=[fill=yellow!30];
\tikzstyle{bpu}+=[fill=blue!20];
\tikzstyle{instruction_fetch}+=[fill=blue!40];
\tikzstyle{instruction_queue}+=[fill=blue!40];
\tikzstyle{waydecode}+=[fill=blue!40];
\tikzstyle{mux}+=[fill=black!10];
\tikzstyle{allocation_queue}+=[fill=blue!40];

\tikzstyle{myarrow} += [->,>=stealth,thick];

\draw[execution_engine] (-1.5,0.75) rectangle +(9.75,-6.5);
\draw (-0.9,-2.5) node [rotate=90] {\Large Execution Engine};

\draw [reorder_buffer] (0,-0.25) rectangle +(8, 0.75) node [pos=.5] {Reorder buffer};
\draw [myarrow] (0.5,-0.25) -- +(0,-0.5) node [pos=.5,right] {\tiny $\mu$OP};
\draw [myarrow] (1.5,-0.25) -- +(0,-0.5) node [pos=.5,right] {\tiny $\mu$OP};
\draw [myarrow] (2.5,-0.25) -- +(0,-0.5) node [pos=.5,right] {\tiny $\mu$OP};
\draw [myarrow] (3.5,-0.25) -- +(0,-0.5) node [pos=.5,right] {\tiny $\mu$OP};
\draw [myarrow] (4.5,-0.25) -- +(0,-0.5) node [pos=.5,right] {\tiny $\mu$OP};
\draw [myarrow] (5.5,-0.25) -- +(0,-0.5) node [pos=.5,right] {\tiny $\mu$OP};
\draw [myarrow] (6.5,-0.25) -- +(0,-0.5) node [pos=.5,right] {\tiny $\mu$OP};
\draw [myarrow] (7.5,-0.25) -- +(0,-0.5) node [pos=.5,right] {\tiny $\mu$OP};

\draw [scheduler] (0,-0.75) rectangle +(8, -0.75) node [pos=.5] {Scheduler};

\draw [execution_units] (0,-2) rectangle +(8, -3) node [yshift=-32.5,xshift=-55,pos=.5] {Execution Units};
\draw [execution_unit] (0.25, -2.25) rectangle +(0.5,-2) node [rotate=90,pos=.5] {\small ALU, AES, \dots};
\draw [execution_unit] (1.25, -2.25) rectangle +(0.5,-2) node [rotate=90,pos=.5] {\small ALU, FMA, \dots};
\draw [execution_unit] (2.25, -2.25) rectangle +(0.5,-2) node [rotate=90,pos=.5] {\small ALU, Vect, \dots};
\draw [execution_unit] (3.25, -2.25) rectangle +(0.5,-2) node [rotate=90,pos=.5] {\small ALU, Branch};
\draw [execution_unit] (4.25, -2.25) rectangle +(0.5,-1.5) node [rotate=90,pos=.5] {\small Load data};
\draw [execution_unit] (5.25, -2.25) rectangle +(0.5,-1.5) node [rotate=90,pos=.5] {\small Load data};
\draw [execution_unit] (6.25, -2.25) rectangle +(0.5,-1.5) node [rotate=90,pos=.5] {\small Store data};
\draw [execution_unit] (7.25, -2.25) rectangle +(0.5,-1.5) node [rotate=90,pos=.5] {\small AGU};

\draw [myarrow] (0.5,-1.5) -- +(0,-0.75) node [pos=.4,right] {\tiny $\mu$OP};
\draw [myarrow] (1.5,-1.5) -- +(0,-0.75) node [pos=.4,right] {\tiny $\mu$OP};
\draw [myarrow] (2.5,-1.5) -- +(0,-0.75) node [pos=.4,right] {\tiny $\mu$OP};
\draw [myarrow] (3.5,-1.5) -- +(0,-0.75) node [pos=.4,right] {\tiny $\mu$OP};
\draw [myarrow] (4.5,-1.5) -- +(0,-0.75) node [pos=.4,right] {\tiny $\mu$OP};
\draw [myarrow] (5.5,-1.5) -- +(0,-0.75) node [pos=.4,right] {\tiny $\mu$OP};
\draw [myarrow] (6.5,-1.5) -- +(0,-0.75) node [pos=.4,right] {\tiny $\mu$OP};
\draw [myarrow] (7.5,-1.5) -- +(0,-0.75) node [pos=.4,right] {\tiny $\mu$OP};

\draw [myarrow,-] (0, -3.55) -- ++(-0.125,0) -- ++(0,0.1+3.75-0.35) -- ++(0.125,0);
\draw [myarrow,-] (0, -3.55-0.125) -- ++(-0.25,0) -- ++(0,0.1+4-0.35) -- ++(0.25,0);
\draw [myarrow,-] (0, -3.55-0.25) -- ++(-0.375,0) -- ++(0,0.1+4.25-0.35) -- ++(0.375,0);
\draw [myarrow,-] (0, -3.55-0.375) -- ++(-0.5,0) -- ++(0,0.1+4.5-0.35) -- ++(0.5,0);

\draw [myarrow,-] (0, -0.875-0.1) -- ++(-0.125,0);
\draw [myarrow,-] (0, -1-0.1) -- ++(-0.25,0);
\draw [myarrow,-] (0, -1.125-0.1) -- ++(-0.375,0);
\draw [myarrow,-] (0, -1.25-0.1) -- ++(-0.5,0);

\draw (-0.9,0.2) node {\footnotesize CDB};

\draw [memory_subsystem] (-1.5,-5.5) rectangle +(9.75, -3);
\draw (-0.75,-7) node [rotate=90, text width=2cm, align=center] {\large Memory\\ Subsystem};

\draw [cache] (0, -6) rectangle +(1.5,-0.5) node [pos=.5] {\footnotesize Load Buffer};
\draw [myarrow,<-] (0.75,-6) -- ++(0, 0.25) -- ++(4.25,0) |- (4.5, -4.25) -- ++(0,0.5);
\draw [myarrow,-] (5,-4.25) |- (5.5, -4.25) -- +(0,0.5);

\draw [myarrow] (0.5,-6.5) -- ++(0,-0.25);
\draw [myarrow] (1,-6.5) -- ++(0,-0.25);

\draw [cache] (1.75, -6) rectangle +(1.5,-0.5) node [pos=.5] {\footnotesize Store Buffer};
\draw [myarrow] (6.5, -3.75) -- ++(0,-2.5) -- ++(-3.25,0);

\draw [myarrow] (2.5,-6.5) -- ++(0,-0.25);

\draw [cache] (0, -6.75) rectangle +(2.75,-1) node [pos=.5] {\small L1 Data Cache};
\draw [tlb] (2.75, -6.75) rectangle +(0.75,-0.5) node [pos=.5] {\footnotesize DTLB};

\draw [myarrow,<->] (3.5,-7) -- ++(0.5,0);
\draw [myarrow,<->] (2.75,-7.5) -- ++(1.25,0);

\draw [tlb] (4,-6.75) rectangle ++(4,-0.5) node [pos=.5] {\footnotesize STLB};
\draw [cache] (4,-7.25) rectangle ++(4,-1) node [pos=.5] {\small L2 Cache};

\begin{scope}[yshift=-0.25cm]
\draw [frontend] (-1.5,1) rectangle +(9.75,6.25);
\draw (-0.9,4) node [rotate=90] {\Large Frontend};

\draw [allocation_queue](0,1.5) rectangle ++(8,0.75) node [pos=.5] {Allocation Queue};
\draw [myarrow] (2.5,1.5) -- +(0,-0.75) node [pos=.4,right] {\tiny $\mu$OP};
\draw [myarrow] (3.5,1.5) -- +(0,-0.75) node [pos=.4,right] {\tiny $\mu$OP};
\draw [myarrow] (4.5,1.5) -- +(0,-0.75) node [pos=.4,right] {\tiny $\mu$OP};
\draw [myarrow] (5.5,1.5) -- +(0,-0.75) node [pos=.4,right] {\tiny $\mu$OP};

\draw [-,mux] (0.25,2.5) -- ++(7.5,0) -- ++(0.25,0.25) -- ++(-8,0) -- ++(0.25,-0.25);
\draw (4,2.5) node [yshift=3] {\tiny MUX};
\draw [myarrow] (4,2.5) -- ++(0,-0.25);

\draw [waydecode] (3,3.25) rectangle ++(5,0.75) node [pos=.5] {\small 4-Way Decode};
\draw [myarrow] (4,3.25) -- +(0,-0.5) node [pos=.5,right] {\tiny $\mu$OP};
\draw [myarrow] (5,3.25) -- +(0,-0.5) node [pos=.5,right] {\tiny $\mu$OP};
\draw [myarrow] (6,3.25) -- +(0,-0.5) node [pos=.5,right] {\tiny $\mu$OP};
\draw [myarrow] (7,3.25) -- +(0,-0.5) node [pos=.5,right] {\tiny $\mu$OP};

\draw [instruction_queue] (3,4) rectangle ++(5,0.75) node [pos=.5] {\small Instruction Queue};

\draw [instruction_fetch] (3,4.75) rectangle ++(5,0.75) node [pos=.5] {\small Instruction Fetch \& PreDecode};

\draw [cache] (0,3.25) rectangle ++(2.5,0.5) node [pos=.5] {\footnotesize $\mu$OP Cache};
\draw [myarrow] (1.25,3.25) -- ++(0,-0.5) node [pos=.5,right] {\tiny $\mu$OPs};

\draw [bpu] (0,4.25) rectangle ++(2.5,1.25) node [pos=.5, text width=2cm, align=center] {Branch\\ Predictor};
\draw [myarrow] (1.25,4.25) -- ++(0,-0.5);

\draw [cache] (3,6) rectangle ++(4,1) node [pos=.5] {L1 Instruction Cache};
\draw [tlb] (7,6.5) rectangle ++(1,0.5) node [pos=.5] {\footnotesize ITLB};

\draw [myarrow] (3,6.5) -- ++(-1.75,0) -- ++(-0,-1);
\draw [myarrow] (5,6) -- ++(0,-0.5);
\end{scope}

\draw [myarrow] (8,-7.75) -- ++(0.75,0) -- ++(0,13.75) -- ++(-1.75,0);
\draw [myarrow] (8,-7) -- ++(0.5,0) -- ++(0,13.5) -- ++(-0.5,0);

\end{tikzpicture}

%% file: images/identity-mapping.tikz
\begin{tikzpicture}

\tikzstyle{page}+=[fill=blue!30];
\tikzstyle{myarrow} += [->,>=stealth,thick];
\draw (0,0) rectangle ++(12,1) node [pos=.5] {\large Physical memory};
\draw (0,1.25) node {0};
\draw (12,1.25) node {max};

\draw [draw=white]  (0,-2) rectangle ++(5.75,1) node [pos=.5] {\large User};
\draw [-] (5.75,-2) -- ++(-5.75,0) -- ++(0,1) -- ++(5.75,0);
\draw (5.75,-2) to[bend right] ++(0,1);
\draw (0,-2.25) node {0};
\draw (5.5,-2.25) node {$2^{47}$};

\draw [draw=white] (6.25,-2) rectangle ++(5.75,1) node [pos=.5] {\large Kernel};
\draw [-] (6.25,-2) -- ++(5.75,0) -- ++(0,1) -- ++(-5.75,0);
\draw (6.25,-2) to[bend right] ++(0,1);
\draw (6.5,-2.25) node {$-2^{47}$};
\draw (12,-2.25) node {$-1$};

\draw [page] (1.25,-2) rectangle ++(0.25,1);
\draw [page] (7.5,-2) rectangle ++(0.25,1);
\draw [page] (2.5,0) rectangle ++(0.25,1);

\draw [-,dashed] (7,-2) -- (7,-1) -- (0,0);
\draw [-,dashed] (8,-2) -- (8,-1) -- (12,0);

\draw [myarrow] (1.375,-1) -- (2.625,0);
\draw [myarrow] (7.625,-1) -- (2.625,0);

\end{tikzpicture}

%% file: images/toy-illustration.tikz
\begin{tikzpicture}

\draw (0,0) rectangle +(2.5,2.8);
\node at (1.25,2.5) {\texttt{$<$instr.$>$}};
\node at (1.25,2) {\texttt{$<$instr.$>$}};
\node at (1.25,1.5) {$\vdots$};
\draw[draw=none,fill=red!20] (0.01,0.01) rectangle +(2.48,1.1);
\node at (1.25,0.8) {\texttt{$<$instr.$>$}};
\node at (1.25,0.3) {\small{\textbf{[ Exception ]}}};
\node[rotate=90] at (2.95, 1.5) {\scriptsize\textsc{executed}};

\draw [draw=none,fill=gray!10] (2.5,-1.6) rectangle ++(1.5,1.6) node[rotate=90,pos=.5] {\parbox{2cm}{\centering \scriptsize \textsc{Executed\\out of\\order}}};
\draw[fill=gray!40] (0,-1.6) rectangle +(2.5,1.6);
\node at (1.25,-.3) {\texttt{$<$instr.$>$}};
\node at (1.25,-0.8) {\texttt{$<$instr.$>$}};
\node at (1.25,-1.3) {\texttt{$<$instr.$>$}};


\node at (-2.5, 1.1) {\textsc{Exception}};
\node at (-2.5, 0.75) {\textsc{Handler}};
\draw[fill=red!5] (-3.75,-1) rectangle +(2.5,1.5);
\node at (-2.5,0.2) {\texttt{$<$instr.$>$}};
\node at (-2.5,-0.3) {\texttt{$<$instr.$>$}};
\node at (-2.5,-0.8) {\small{\textbf{[ Terminate ]}}};

\draw (0,0.25) edge[,->,thick,>=stealth] (-1.25,0.25);
\end{tikzpicture}

%% file: images/fr-toy.tikz
\begin{tikzpicture}
\begin{axis}[
mlineplot,
style={font=\footnotesize},
xlabel={Page},
ylabel={Access time [cycles]},
ylabel style={text width=2cm,align=center},
width=0.95\hsize,
scaled y ticks=false,
xmin=0,
xmax=255,
height=3cm
]
\addplot+[blue,thick,mark options={draw=blue,fill=blue},mark=*,mark size=0.5] table[x=data,y=time,col sep=space] {data/fr-toy.csv};

\end{axis}
\end{tikzpicture}

%% file: images/building-blocks.tikz
\begin{tikzpicture}

\draw[postaction={draw,pattern=north west lines, pattern color=red!10}] (-0.85,-1.25) rectangle +(3.7,3.25);
\node at (1.05,1.72) {\small Exception Handling/};
\node at (1.05,1.35) {\small Suppression};

\draw[fill=white] (-0.5,-1) rectangle +(3, 2);
\node at (1,0.75) {Transient};
\node at (1,0.25) {Instructions};

\draw[fill=white] (4,0) rectangle +(2, 1);
\node at (5,0.5) {Secret \includegraphics[height=0.7em]{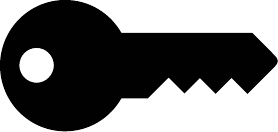}};

\draw[fill=yellow!10,densely dotted] (-0.4,-0.9) rectangle +(2.8, 0.8);
\node at (1,-0.3) {\scriptsize Microarchitectural};
\node at (1,-0.7) {\scriptsize State Change};

\draw[dashed] (-1.1,-1.8) rectangle +(7.5,4);
\node at (-0.1,-1.55) {\scriptsize Section 4.1};

\draw[fill=yellow!10] (-0.5,-3.5) rectangle +(3, 1);
\node at (1,-2.8) {Architectural};
\node at (1,-3.2) {State};
\draw[>=stealth,->,thick] (1,-0.9)  to node[near end,above,yshift=-1em,xshift=1.25em]{\scriptsize Transfer (Covert Channel)} (1,-2.5);
\draw[dashed] (-1.1,-3.95) rectangle +(7.5,1.95);

\draw (4,-3.5) rectangle +(2, 1);
\node at (4.95,-2.75) {Recovered};
\node at (4.62,-3.2) {Secret};
\node at (5.55,-3.2) {\includegraphics[height=0.7em]{images/key.pdf}};
\draw[>=stealth,->] (2.5,-3) to node[midway,sloped,above]{\scriptsize Recovery} (4,-3);

\draw[>=stealth,->,dotted,thick] (5,0) to node[midway,sloped,above]{\scriptsize Leaked} (5,-2.5);

\draw[>=stealth,<-] (2.5,0.5) to node[midway,sloped,above,xshift=0.5em]{\scriptsize Accessed} (4,0.5);
\node at (-0.1,-3.75) {\scriptsize Section 4.2};

\end{tikzpicture}